# Dual-gated bilayer graphene hot electron bolometer

J. Yan[1,2], M.-H. Kim[1,2], J.A. Elle[2,3], A.B. Sushkov[1,2], G.S. Jenkins[1,2],

H.M. Milchberg[2,3], M.S. Fuhrer[1,2]*, and H.D. Drew[1,2]

[1] Center for Nanophysics and Advanced Materials, [2] Department of Physics,

[3] Institute for Research in Electronics and Applied Physics,

University of Maryland, College Park, Maryland 20742, USA

*email: mfuhrer@umd.edu

**Detection of infrared light is central to diverse applications in security, medicine, astronomy, materials science, and biology. Often different materials and detection mechanisms are employed to optimize performance in different spectral ranges. Graphene is a unique material with strong, nearly frequency-independent light-matter interaction from far infrared to ultraviolet, with potential for broadband photonics applications. Moreover, graphene's small electron-phonon coupling suggests that hot-electron effects may be exploited at relatively high temperatures for fast and highly sensitive detectors in which light energy heats only the small-specific-heat electronic system. Here we demonstrate such a hot-electron bolometer using bilayer graphene that is dual-gated to create a tunable bandgap and electron-temperature-dependent conductivity. The measured large electron-phonon heat resistance is in good agreement with theoretical estimates in magnitude and temperature dependence, and enables our graphene bolometer operating at a temperature of 5 K to have a low noise equivalent power (33 fW/Hz$^{1/2}$). We employ a pump-probe technique to directly measure the intrinsic speed of our device, >1 GHz at 10 K.**

Graphene, a two-dimensional (2D) hexagonal lattice of carbon, has remarkable electronic, thermal, and mechanical properties[1]. Graphene's exceptional stiffness[2] leads to exceptionally low electron phonon interaction and high intrinsic mobility[3-6]. Graphene's unique band structure[7] results in charge carriers which obey the massless Dirac equation in 2D[8,9] and gives rise to a quantized absorption coefficient which is frequency independent over a broad range from far infrared to ultraviolet[10,11]. In this article, we exploit the small electron-phonon scattering[3] and broadband photon absorption [10,11] to realize a sensitive broadband photon detector with graphene. We use infrared optical excitations to generate hot carriers[12,13] in electric-field tuned graphene bilayers[14-16] which exhibit a bandgap[16-18].

There are previous efforts to use graphene as photo detectors, most notably the studies of rectifying p-n junctions[12,19], graphene/metal junctions [20-22], and graphene monolayer/bilayer interfaces[23]. The underlying mechanisms have been attributed to photovoltaic and/or thermoelectric effects. The hot electron bolometer (HEB) we present here is conceptually different and it makes use of the 2D graphene bulk instead of 1D rectifying interfaces. Our photo detector utilizes thermal decoupling of the electrons from the lattice[12]. The small electronic specific heat allows for fast response times, high sensitivity, and low noise equivalent power. Even the unoptimized detector presented here already has similar sensitivity and much lower intrinsic noise than commercial silicon bolometers and superconducting transition edge sensors[24] operating at similar temperatures (5 K), and the intrinsic speed of our device is three to five orders of magnitude faster. Furthermore, the strong temperature dependence of the heat resistance, specific heat and electron scattering time suggests much higher sensitivity and lower noise at lower temperatures (less than 1 K) which could open doors for engineering graphene-based single photon detectors which are of great importance for contemporary astronomy[25].

The essential ingredients of our graphene HEB are the weak electron-phonon coupling in graphene[3], and the temperature dependent resistance induced by the insulating state in dual-gated bilayer graphene (DGBLG) [14-16]. When light is absorbed by DGBLG, the electrons heat up easily due to their small specific heat. The weak electron-phonon interaction creates a bottleneck in the heat path, decoupling electrons thermally from the phonon bath. Since the charge transport properties of insulating DGBLG are sensitive to electron temperature, light illumination causes a resistance change $\Delta R$ in the sample. This photon absorption induced $\Delta R$ is then converted to a detectable electrical signal.

**Dual-gated bilayer graphene device and photoresponse**

Figure 1a shows schematically our device structure together with the gating scheme. Figure 1b shows the optical image of one of the samples (see Methods for details of device fabrication and measurement). Figure 1c shows a typical optical response of DGBLG as a function of dc bias current $I_{dc}$ under illumination by continuous wave (cw) $CO_2$ laser light (wavelength 10.6 μm) chopped at 700 Hz. The signal $\Delta V = I_{dc}\Delta R$ is detected with a lock-in amplifier. Similar observations are made with other light sources, such as a 2 μm cw laser, a 1.03 μm pulsed laser, and a broadband Globar mid infrared illumination. The signal in Fig.1c increases with $I_{dc}$ and reaches more than 0.7mV at high currents. Assuming 4% of light absorption by DGBLG[10,11] and taking into account effects due to the silicon substrate and the Nichrome top gate, we estimate the absorbed power to be 3.7 nW (0.45% of incident light); this corresponds to a voltage responsivity of about $2\times10^5$ V/W. As a comparison, commercial silicon bolometers have responsivities between $10^4$ and $10^7$ V/W.

**Origin of the photoresponse**

We now discuss the origin of the photoresponse, considering the possibilities that it may be either photoconductive or bolometric. In the first case, photons create electron-hole pairs providing additional conductance channels in the device. The conductance change is given by $\Delta\sigma = \Delta n e \mu$ where $\Delta n$ is the steady-state photo-excited charge density determined by the incident laser power and carrier lifetime, $e$ is electron charge, $\mu$ is mobility. This leads to a voltage

$$\Delta V = I_{dc}\Delta R = I_{dc}\frac{L}{W}\Delta\left(\frac{1}{\sigma}\right) = I_{dc}\frac{W}{L}R^2\Delta\sigma \quad , \tag{1}$$

where $L/W$ is determined by the aspect ratio of the device. In the second case, light absorption creates heat and a temperature change $\Delta T$; the signal is given by

$$\Delta V = I_{dc}\Delta R = I_{dc}\frac{dR}{dT}\Delta T \tag{2}.$$

We note that in Fig.1c the saturation behavior and the decreasing of $\Delta R$ with $I_{dc}$ already suggest that modest power dissipation in the sample may cause significant heating. For $I_{dc} = 150\text{nA}$, the electric power dissipated is 3.6nW (sample resistance 160KΩ at 5 K), comparable to the absorbed light power. The significant decrease of $\Delta R$ with increasing $I_{dc}$ suggests that the sample is heated, causing the response of the device to diminish due to higher temperatures.

The combination of optical and transport methods provides a powerful tool to determine whether the optical signal is created by a photoconductive or a bolometric mechanism. We have measured the two-probe resistance $R$ and photo response voltage $\Delta V$ of another DGBLG device as a function of top gate and back gate voltages $V_{tg}$ and $V_{bg}$. Figures 2a and 2b show 2D maps of $R(V_{tg},V_{bg})$ and $\Delta V(V_{tg},V_{bg})$ at a temperature of 6 K. A check of $\Delta V$ for the same ($V_{tg}, V_{bg}$) before and after 2D mapping indicated the laser output was stable within 10%.

The $R(V_{tg}, V_{bg})$ map (Fig. 2a) reproduces previous observations for DGBLG[14-16]. The top and back gates produce displacement fields $D_t = -\varepsilon_t(V_{tg}-V_{tg0})/d_t$ and $D_b = \varepsilon_b(V_{bg}-V_{bg0})/d_b$ where $V_{tg0}(V_{bg0})$ are charge neutrality point shifts due to unintentional chemical doping above (below) the graphene, and $\varepsilon_t(\varepsilon_b)$ and $d_t(d_b)$ are the dielectric constants and thicknesses of the top (back) gates, respectively. For the device here, $V_{tg0} \approx 42$V, $V_{bg0} \approx 7$V, $\varepsilon_t \approx \varepsilon_b \approx 3.9$, $d_t \approx d_b \approx 300$nm. The average displacement field $\overline{D} = (D_t+D_b)/2$ across the DGBLG opens a bandgap at the charge neutrality point[17], while the difference in displacement fields $D_t - D_b$ produces doping. Figure 2c shows cuts of $R$ at various $\overline{D}$. $R$ is peaked when DGBLG is charge neutral at $D_t = D_b$ and decreases as the sample is doped with holes ($D_t > D_b$) or electrons ($D_t < D_b$).

The photo response voltage $\Delta V(V_{tg}, V_{bg})$ (Fig. 2b and the cuts in Fig.2d) displays qualitatively similar behavior to $R(V_{tg}, V_{bg})$: $\Delta V$ is largest at charge neutrality, increases with increasing $\overline{D}$, and decreases with doping. However it is clear from Figs. 2c and 2d that the photo response is much more strongly peaked about charge neutrality, and varies more strongly with bandgap. If we assume that the mobility and carrier lifetime do not vary significantly with the gate voltages, it is expected from equation (1) that $\Delta V \propto R^2$. In Fig. 2e we compare $\Delta V$ and $R^2$ along the charge neutral cut with varying $\overline{D}$. We observe that $\Delta V$ varies much more strongly than $R^2$ suggesting that the origin of the observed photo response is not photoconduction. Cuts along constant $\overline{D}$ with varying carrier density (Fig.2f) give the same conclusion.

**Electrical measurement of the thermal resistance**

The remaining possibility is that the signal is bolometric. In this case, we expect to observe similar resistance changes in our device when heated electrically rather than optically. Since the electrical power is easier to characterize, the thermal resistance of the electron system to the heat bath can be quantitatively extracted. To probe the effect of electrical heating, we pass both dc and ac currents in DGBLG, and detect the voltage at the fundamental ac frequency as well as its second harmonic. In the limit of small temperature rise due to heating, the first harmonic signal gives the resistance $R$ of the sample as a function of temperature $T$ and $I_{dc}$ shown in Fig.3a. The decrease of $R$ with higher $T$ and $I_{dc}$ reflects the insulating transport behavior of DGBLG. The second harmonic signal is given by (see Methods)

$$\Delta V(2\omega) = \frac{3}{\sqrt{2}} I_{dc} I_{ac}^2 \frac{dR}{dP} \qquad (3),$$

where $\Delta V(2\omega)$ and $I_{ac}$ (40nA) are root mean square values. In fig.3b the signal changes sign with $I_{dc}$ and is linear when $I_{dc}$ is small, consistent with the description of Eq.(3). At higher $I_{dc}$ (higher temperature rise) the signal saturates, similar to the saturation behavior in Fig.1c; assuming that the $I_{dc}$ dependence of $R$ is due to heating, non-linearity of $\Delta V(2\omega)$ at high $I_{dc}$ reflects that $dR/dP$ decreases with temperature.

Using equation (3) we convert the data in Fig.3b to $dR/dP$ as a function of $I_{dc}$ at various $T$. With $I_{dc}$ dependence of $R$ in Fig. 3a, we plot in Fig.3c $dR/dP$ vs. $R$, and find that $dR/dP$ is a unique function of $R$. This indicates that $R$ is a function of $T$ alone, and $dR/dP$

= *(dR/dT)(dT/dP)*; i.e. *R* does not depend explicitly on *P* which might occur due to other non-linearities in the system.

In Fig.3c *dR/dP* ranges from one to tens of KΩ/nW, consistent with the typical resistance changes observed in optical measurements as shown in Fig.1c. Considering that the optical power absorption estimation is quite rough, in the following we rely on the electric Joule heating measurements to quantify physical parameters linked to the BLG hot electron bolometer.

A key quantity of interest in the bolometric effect is the thermal resistance $R^h = dT/dP$, or equivalently thermal conductance $G^h = dP/dT$. Having measured *dR/dP* as well as the temperature dependence of the resistance *R(T)*, we can now calculate *dT/dP* directly from the transport data using Joule heating. In Fig.s 4a and 4b we phenomenologically fit *R* and *dR/dP* with power law temperature dependences: $R = 67.5 \times (T/5)^{-0.75}$ KΩ, $-dR/dP = 20 \times (T/5)^{-5.2}$ KΩ/nW. This allows us to obtain the heat resistance

$$R^h = \frac{dT}{dP} = \frac{dR/dP}{dR/dT} = 2 \times \left(\frac{T}{5}\right)^{-3.45} \text{K/nW} \qquad (4).$$

In Figure 4c we use the experimental *R(T)* to replot the data from Fig. 3c as $R^h$ vs. *T*. The experimental data (symbols) are well described by equation (4) (red line).

**Discussion of thermal conductance pathways in the device**

We now discuss the origin of the observed thermal resistance $R^h$. Figure 4d shows the heat diagram of our device. There are two parallel channels for heat transfer from graphene to the bath: through the $SiO_2$ gate dielectrics, or through the electrical contact leads. Using the thermal conductivity of amorphous $SiO_2$ of $10^{-3}$ W/cm-K at 5 K[26], we estimate a thermal resistance of the order $10^{-5}$ K/nW for our device geometry. The graphene/$SiO_2$ thermal contact resistance is $2\times10^{-8}$ m$^2$K/W at 42 K[27]; assuming $T^3$ extrapolation, at 5 K for our device $R^h_{BLG/SiO2} \approx 10^{-4}$ K/nW. Since both $R^h_{SiO2}$ and $R^h_{BLG/SiO2}$ are much smaller than our observed thermal resistance, the relaxation of hot carriers with the graphene lattice characterized by $R^h_{el-ph}$ must dominate this heat flow channel. From the electrical contact resistance of order 10 kΩ or larger and using the Wiedemann–Franz law, we estimate the thermal resistance of the electrical contacts to be $R^h_{BLG/Au} > 80$ K/nW at 5 K, well above our observed heat resistance suggesting that the hot electrons are phonon cooled instead of diffusively cooled.

The analysis above indicates that the bottleneck in the heat path is $R^h_{el-ph}$ between electrons and phonons in DGBLG. The heat resistance shown in Fig.4c is thus a quantitative characterization of the intrinsic weak electron-phonon interaction in graphene. The temperature dependence of $R^h$ was calculated in Ref. 28. It was shown that for $T \gg T_{BG}$ where $T_{BG}$ is the Bloch-Grüneisen temperature, above which phonons of electron Fermi wavevector $k_F$ are in the equipartion regime, $R^h \propto T^{-1}$. For $T \ll T_{BG}$, the bosonic nature of these phonons is manifested and $R^h \propto T^{-3}$. While our sample is nominally charge neutral, it is widely accepted that disorder creates electron-hole puddles[29] and thus $T_{BG}$ is nonzero. Assuming a disorder-induced charge density of about $10^{12}$ cm$^{-2}$ [30] and a sound velocity $v_s = 2.6\times10^4$ m/s [31], we find $T_{BG} \approx 2\hbar v_s k_F/k_B \approx 70$ K where $k_B$ is the Boltzmann constant. The power-law exponent -3.45 in equation (4) reasonably agrees with the expected exponent -3. Transport measurements show that the Bloch-Grüneisen regime behavior occurs for $T < 0.2 T_{BG}$ [31] consistent with our observation of approximately $T^3$ dependence of $R^h$ below 10K.

We can make a quantitative estimate of the magnitude of $R^h$ from Ref. 28, which gives

$$R^h = \frac{15\rho\hbar^5 v_F^3 v_s^3}{\pi^2 D^2 \gamma_1 k_B^4} \sqrt{\frac{|E_F|}{\gamma_1}} \frac{T^{-3}}{A} \qquad (5).$$

Using a deformation potential $D = 18$ eV [3], interlayer coupling $\gamma_1 = 390$ meV [32], bilayer graphene mass density $\rho = 1.5\times10^{-6}$ kg/m$^2$, monolayer graphene Fermi velocity $v_F = 10^6$ m/s, sound velocity $v_s = 2.6\times10^4$ m/s [31], sample Fermi energy $E_F = 33$ meV, and sample area $A = 100$ μm$^2$, we find the calculated thermal resistance to be $0.6 \times (T/5)^{-3}$ nW/K. The agreement to our observation of $2 \times (T/5)^{-3.45}$ nW/K is very good considering the roughness of carrier density estimation and various approximations used in deriving equation (5) [28].

**Response speed of the graphene hot electron bolometer**

Another important bolometer parameter is its response time given approximately by the ratio of the electronic heat capacity to thermal conductance. The electron specific heat is $C = (\pi^2/3)\upsilon(E_F)k_B^2 T$ where $\upsilon(E_F) \approx \gamma_1/(\pi\hbar^2 v_F^2)$, the density of states for bilayer graphene. For our sample the heat capacity is estimated to be $1.12\times10^{-19}$ J/K at 5 K. The response time is then expected to be $\tau \approx R^h C \approx 0.2\times(T/5)^{-2.45}$ ns. This fast response is beyond the detection limit of our present circuit, and indeed we found that our signal is independent of frequency up to $10^4$ Hz, the RC time constant of the circuit limited by the large capacitance of the electrode contact pads to the gate.

Hot electron energy relaxation dynamics in graphene has been studied by time-resolved photoluminescence[33], transmission[34], reflection[35] and Raman spectroscopy[36]. However all these techniques are sensitive only to electron temperatures hundreds to thousands of Kelvin above the lattice temperature. In contrast, in the 2$^{nd}$ harmonic transport measurements discussed in the previous section, the electrons are only a few or a fraction of one Kelvin above the lattice.

To measure the intrinsic speed of our HEB, we make use of the non-linear nature of the photoresponse and study it with a 1.03 μm pulsed laser. In the set up (details in Methods), graphene absorbs an average of 0.13nW from the pump and probe pulses, and the probe delay time (with respect to the pump pulse) $t_D$ is tunable from -1.5 to 2.5 ns. We measure the signal $\Delta V$ of equation (2) due to the pump beam $\Delta V_{pu}$, the probe beam $\Delta V_{pr}$ and the pump+probe beam $\Delta V_{pp}$ as a function of $t_D$. The ratio $r = \Delta V_{pp}/(\Delta V_{pu}+\Delta V_{pr})$ reflects the non-linear power dependence of the photoresponse.

As shown in Fig.5, $r$ approaches 1 for large $t_D$, and decreases to a minimum but non-zero value as $t_D$ goes to 0. This can be understood from the strong temperature dependence of d$R$/d$P$ in Fig.4b. Upon interaction with the first pulse the electron temperature rises and the resistance decreases. If the second pulse arrives before the electrons reach thermal equilibrium with the lattice, the additional resistance change is not as large since d$R$/d$P$ is smaller at higher electron temperatures.

In the two traces in Fig.5, the time constants are about 0.25 ns at 4.55 K and 0.1 ns at 10 K, in good agreement with our estimations above. We note that since we used photons of relatively high energy in this measurement, relaxation processes involving optical phonons should also occur on the picoseconds timescale[33-36], not detected in our measurement. The observed long relaxation times confirm that we are studying a different regime of hot carrier relaxation and mark the first time-resolved measurement of electron-acoustic phonon scattering in graphene.

**Noise equivalent power**

We now discuss the noise equivalent power (NEP) which characterizes the signal-to-noise ratio of photodectectors. We consider two noise sources, thermal noise and Johnson–Nyquist noise. While in a real measurement other noise sources such as those from the amplifier could also contribute[37], these two are intrinsic to the bolometer itself. The thermal fluctuation NEP is given by $\sqrt{4k_B T^2 / R^h}$ which at 5K is 0.26fW/Hz$^{1/2}$ for our measured $R^h$ of 2 K/nW. The Johnson–Nyquist noise contribution to NEP is given by[37] the Johnson noise $\sqrt{4k_B TR} = 6.7 nV/Hz^{1/2}$ divided by the voltage responsivity $2\times10^5$ V/W which gives 33 fW/Hz$^{1/2}$ at 5 K. These values are exceptionally small as compared with the 200-2000 fW/Hz$^{1/2}$ NEP of commercially-available silicon bolometers. Our graphene HEB is also competitive with superconducting transition edge sensors working at 4.2 K where response time and NEP were found to be 1.2 ms and 120 fW/Hz$^{1/2}$ [24].

At lower temperatures, we expect significant improvements in device performance. Assuming $T^3$ dependence of $R^h$, and a linear T dependence of the specific heat we deduce a thermal NEP of $1.5\times10^{-20}$ W/Hz$^{1/2}$ at 100 mK with a response time of 3 μs. Further improvements are expected with smaller devices achievable with current lithography technology. This would outperform the sensitivity limit of micromachined bolometers due to the 1D quantized thermal conductance[38] and approach the sky background NEP of ~$10^{-20}$ W/Hz$^{1/2}$ in the terahertz range at reasonable device temperatures (300 mK for a device of 1 $\mu m^2$). These fundamental parameters are very promising compared with the state-of-the-art superconducting bolometers[39]. In our present version of a graphene HEB the Johnson-Nyquist noise is greater than the phonon noise. However, other configurations may overcome this limitation by decreasing impedance and increasing the responsivity of the device.

**Conclusions**

We have demonstrated a hot electron bolometer based on insulating dual-gated bilayer graphene. Our graphene bolometer has performance characteristics which are favorable to commercial silicon bolometers, and competitive with those of superconducting transition-edge detectors. The extremely small electron-phonon coupling in graphene implies that properly-designed graphene-based hot electron bolometers may operate to significantly higher temperatures. The small thermal NEP makes it a promising candidate for single photon detectors for astronomers. There is significant room for improvement. The device can be made smaller, decreasing its heat capacity and increasing the heat resistance, improving both NEP and sensitivity. If more insulating graphene could be used, the responsivity may be improved due to stronger temperature dependence of the resistance.

We also note two challenges for the graphene bolometer. The characteristic electrical impedance ($h/2e^2$ = 13 k$\Omega$) of a two-dimensional conductor such as graphene is high compared to the impedance of free space ($2h/\alpha e^2$ = 377 $\Omega$) indicating that graphene-antenna coupling and graphene-amplifier coupling may be difficult at high frequencies. In addition, graphene's optical conductivity of $\pi e^2/2h$ per layer gives an absorption of $\pi\alpha$ = 2.3% per layer, making it an inefficient absorber. Both of these challenges could be addressed by using (suitably insulating) multi-layer graphene to reduce the dc and optical impedances.

**Methods**

The bilayer graphene sample (area 100 $\mu m^2$) is deposited on a conducting silicon substrate with 300nm thick $SiO_2$ by mechanical exfoliation of natural graphite (Nacional de Grafite Ltd.a.). Electrical contacts are made with standard electron-beam lithography followed by thermal evaporation of chromium/gold (5/100 nm). We protect the graphene flake with electron-beam exposed hydrogen silsesquioxane before sputtering $SiO_2$ as the top gate dielectric. The total thickness of hydrogen silsesquioxane and sputtered $SiO_2$ is close to 300nm, so the gating efficiencies of top and back gates are similar. Thin Nichrome (sheet resistivity 200 $\Omega$) is used as a semi-transparent top gate.

Carrier density and band gap in bilayer graphene are tuned with two Keithley 2400 source meters. We used both dc and ac currents to bias the sample. The dc bias is provided with either a battery or another Keithley source meter, and the ac excitation is from a Stanford SR830 lock-in amplifier. The transport characterization in Fig.2a is done with a 10 nA low frequency ac bias. For the electrical measurement of heat resistance in Fig. 3, both dc and ac biases ($I = I_{dc} + \sqrt{2} I_{ac} \cos \omega t$) are applied. The voltage drop on the sample is given by $V = I \left( R + \frac{dR}{dP} I^2 R \right)$, taking into account electrical Joule heating. The second harmonic signal is then given by $\Delta V(2\omega) = \frac{3}{\sqrt{2}} I_{dc} I_{ac}^2 \frac{dR}{dP}$. In the measurements, the first and second harmonic signals are recorded simultaneously at different temperatures to obtain resistance dependence of *dR/dP*.

We used various light sources to study photo response of the sample, a 10.6 $\mu$m $CO_2$ laser, a 2 $\mu$m cw laser, a 1.03 $\mu$m pulsed laser, and a broadband Globar mid infrared illumination. In the paper, we show data collected with the $CO_2$ laser (Fig.1c and Fig.2b) and with the pulsed laser (Fig.5). In these measurements we use a battery box to apply a dc bias across a 1 M$\Omega$ resistor in series with the sample. The light is chopped at a few hundred Hertz and the signal is detected with a lock-in amplifier. For the pump-probe measurement, we use a high-power pulsed fiber laser operating at 1.03 $\mu$m (PolarOnyx Uranus 05-600-INS). Ten pico-second pulses are generated at a repetition-rate of 38 MHz (13 ns pulse separation). Each pulse is split into two identical pulses, with energy of ~ $10^{-18}$ J/$\mu m^2$. An optical delay line is used to impose a time delay $t_D$ on one of them (probe pulse). The range of $t_D$ is from -1.5 to 2.5 ns. Graphene absorbs an average power of 0.13 nW from the pulsed laser. For each $t_D$, the photoresponse voltage is measured for the pump, the probe, and the pump+probe pulses respectively.

**References:**

1. Castro Neto, A. H., Guinea, F., Peres, N. M. R., Novoselov, K. S. & Geim, A. K.

The electronic properties of graphene. *Rev. Mod. Phys.* **81**, 109–162 (2009).

2. Lee, C., Wei, X., Kysar, J.W. & Hone, J. Measurement of the elastic properties and intrinsic strength of monolayer graphene. *Science* **321**, 385-388 (2008).

3. Chen, J.-H., Jang, C., Xiao, S., Ishigami, M. & Fuhrer, M. S. Intrinsic and extrinsic performance limits of graphene devices on $SiO_2$. *Nature Nanotech.* **3**, 206–209 (2008).

4. Bolotin, K. I. *et al*. Ultrahigh electron mobility in suspended graphene. *Solid State Commun.* **146**, 351–355 (2008).

5. Du, X., Skachko, I., Barker, A. & Andrei, E. Y. Approaching ballistic transport in suspended graphene. *Nature Nanotech.* **3**, 491–495 (2008).

6. Dean, C. R. *et al*. Boron nitride substrates for high-quality graphene electronics. *Nature Nanotech.* **5**, 722–726 (2010).

7. Wallace, P. R. The band theory of graphite. *Phys. Rev.* **71**, 622-634 (1947).

8. Novoselov, K. *et al*. Two-dimensional gas of massless Dirac fermions in graphene. *Nature* **438**, 197-200 (2005).

9. Zhang, Y., Tan, Y., Stormer, H. & Kim, P. Experimental observation of the quantum Hall effect and Berry's phase in graphene. *Nature* **438**, 201-204 (2005).

10. Nair, R. R. *et al.* Fine structure constant defines visual transparency of graphene. *Science* **320**, 1308 (2008).

11. Mak, K.F., Sfeirb, M.Y., Misewich, J.A., & Heinz, T. F. The evolution of electronic structure in few-layer graphene revealed by optical spectroscopy. *Proc. Natl Acad. Sci. USA* **107**, 14999-15004 (2010).

12. Gabor, N.M. *et al*. Hot carrier–assisted intrinsic photoresponse in graphene.

*Science* DOI: 10.1126/science.1211384 (2011).

13. Kalugin, N.G. *et al*. Graphene-based quantum Hall effect infrared photodetector operating at liquid nitrogen temperatures. *Appl. Phys. Lett*. **99**, 013504 (2011).

14. Oostinga, J. B., Heersche, H. B., Liu, X. L., Morpurgo, A. F. & Vandersypen, L. M. K. Gate-induced insulating state in bilayer graphene devices. *Nature Mater*. **7**, 151–157 (2008).

15. Zhou, K. & Zhu, J. Transport in gapped bilayer graphene: The role of potential fluctuations. *Phys. Rev. B* **82**, 081407(R) (2010).

16. Yan, J. & Fuhrer, M.S. Charge transport in dual gated bilayer graphene with Corbino geometry. *Nano Lett.* **10**, 4521-4525 (2010).

17. Zhang, Y. *et al*. Direct observation of a widely tunable bandgap in bilayer graphene. *Nature* **459**, 820-823 (2009).

18. Mak, K.F., Lui, C.H., Shan, J. & Heinz, T.F. Observation of an electric-field-induced band gap in bilayer graphene by infrared spectroscopy. *Phys. Rev. Lett*. **102**, 256405 (2009).

19. Lemme, M.C., *et al*. Gate-activated photoresponse in a graphene p-n Junction. *Nano Lett.* **11**, 4134–4137 (2011).

20. Park, J., Ahn, Y. H., & Ruiz-Vargas, C. Imaging of photocurrent generation and collection in single-layer graphene. *Nano Lett*. **9**, 1742-1746 (2009).

21. Xia, F. N., Mueller, T., Lin, Y. M., Valdes-Garcia, A. & Avouris, P. Ultrafast graphene photodetector. *Nature Nanotech.* **4**, 839–843 (2009).

22. Mueller, T., Xia, F. & Avouris, P. Graphene photodetectors for high-speed optical communications. *Nature Photon.* **4**, 297–301 (2010).

23. Xu, X. D., Gabor, N. M., Alden, J. S., van der Zande, A. M. & McEuen, P. L. Photo-thermoelectric effect at a graphene interface junction. *Nano Lett.* **10**, 562-566 (2010).

24. Skidmore, J. T., Gildemeister, J., Lee, A. T., Myers, M. J. & Richards, P. L. Superconducting bolometer for far-infrared Fourier transform spectroscopy. *Appl. Phys. Lett*. **82**, 469 (2003).

25. Richards, P.L. & McCreight, C.R. Infrared detectors for astrophysics. *Phys. Today* **58**, 41-47 (2005).

26. Stephens, R.B. Low-temperature specific heat and thermal conductivity of noncrystalline dielectric solids. *Phys. Rev. B* **8**, 2896–2905 (1973).

27. Chen, Z., Jang, W., Bao, W., Lau, C.N. & Dames, C. Thermal contact resistance between graphene and silicon dioxide. *Appl. Phys. Lett*. **95**, 161910 (2009).

28. Viljas, J.K. & Heikkilä T.T. Electron-phonon heat transfer in monolayer and bilayer graphene. *Phys. Rev. B* **81**, 245404 (2010).

29. Martin, J. *et al*. Observation of electron–hole puddles in graphene using a scanning single electron transistor. *Nature Phys*. **4**, 144–148 (2008).

30. Yan, J., Henriksen, E. A., Kim, P. & Pinczuk, A. Observation of anomalous phonon softening in bilayer graphene. *Phys. Rev. Lett*. **101**, 136804 (2008).

31. Efetov, D.K. & Kim, P. Controlling electron-phonon interactions in graphene at ultrahigh carrier densities. *Phys. Rev. Lett*. **105**, 256805 (2010).

32. Zhang, L. M. *et al*. Determination of the electronic structure of bilayer graphene from infrared spectroscopy. *Phys. Rev. B* **78**, 235408 (2008).

33. Lui, C.H., Mak, K.F., Shan, J. & Heinz, T.F. Ultrafast Photoluminescence from

Graphene. *Phys. Rev. Lett*. **105**, 127404 (2010).

34. Wang, H.N. *et al*. Ultrafast relaxation dynamics of hot optical phonons in graphene. *Appl. Phys. Lett*. **96**, 081917 (2010).

35. Hale, P. J., Hornett, S. M., Moger, J., Horsell, D. W. & Hendry, E. Hot phonon decay in supported and suspended exfoliated graphene. *Phys. Rev. B* **83**, 121404(R) (2011).

36. Chatzakis, I., Yan, H., Song, D., Berciaud, S. & Heinz, T.F. Temperature dependence of the anharmonic decay of optical phonons in carbon nanotubes and graphite. *Phys. Rev. B* **83**, 205411 (2011).

37. Richards, P. L. Bolometers for infrared and millimeter waves. *J. Appl. Phys*. **76**, 1–24 (1994).

38. Schwab, K., Henriksen, E. A., Worlock, J. M. & Roukes, M. L. Measurement of the quantum of thermal conductance. *Nature* **404**, 974–977 (2000).

39. Wei, J. *et al*. Ultrasensitive hot-electron nanobolometers for terahertz astrophysics. *Nature Nanotech*. **3**, 496–500 (2008).

**Acknowledgements**

We thank John Melngailis and Tony F. Heinz for discussions. This work is supported by IARPA, the ONR MURI program, and NSF grants DMR-0804976 and DMR-1105224. J.A.E. and H.M.M. acknowledge the support of NSF.

**Author contributions**

M.S.F. and H.D.D. conceived the project. J.Y. fabricated the devices and performed the transport measurements. M.H.K, J.Y., A.B.S. and G.S.J. conducted the photo response experiments. J.A.E. and H.M.M. assisted in the pump-probe measurements. J.Y., M.H.K, M.S.F. and H.D.D. analyzed data and wrote the paper. All authors discussed and contributed to writing the paper.

**Competition financial interests**

The authors have no competition financial interests to declare.

**Figure legends**

**Figure 1: Bilayer graphene device and typical optoelectronic photoresponse. a**, Schematic illustration of device geometry and electric field effect gating. **b**, Optical micrograph of a bilayer graphene device. The scale bar is 5μm. For clarity the image shows the sample prior to top gate dielectric and metal deposition. **c**, Photo response of dual-gated bilayer graphene. The sample is gated to charge neutral position with a displacement field of 0.45 V/nm. The blue squares are the photo response $\Delta V$. The red circles are the electrical resistance change $\Delta R = \Delta V/I_{dc}$.

**Figure 2: Dependences of $R$ and $\Delta V$ on $V_{tg}$ and $V_{bg}$. a**, Two-dimensional map of $R(V_{tg},V_{bg})$ measured in the dark. **b**, Two-dimensional map of $\Delta V(V_{tg},V_{bg})$ with dc bias of 255 nA. **c,d**, One-dimensional cuts of $R$ and $\Delta V$ through data in panels **a** and **b**. The two black cuts correspond to varying the average displacement field (varying the bandgap) while maintaining charge neutrality ($n = 0$). Others correspond to varying the charge density at fixed average displacement fields (fixed bandgap). **e**, Comparison of normalized $R^2$ and $\Delta V$ with varying average displacement fields at charge neutrality, where $R_{\bar{D}=0} = 17K\Omega$, $\Delta V_{\bar{D}=0} = 5.4\mu V$. **f**, Comparison of normalized $R^2$ and $\Delta V$ with varying charge density at a fixed average displacement field of 0.45 V/nm, where $R_{n=0} = 30K\Omega$, $\Delta V_{n=0} = 28\mu V$. The measurements are done at a temperature of 6 K in cold helium gas.

**Figure 3: Electrical heating of DGBLG at charge neutrality with $\bar{D}$ = 0.55 V/nm. a**, Dependence of resistance $R$ on dc current $I_{dc}$ at various temperatures. **b**, Dependence of second harmonic signal on $I_{dc}$ at various temperatures. **c**, Data from **a** and **b** replotted as $dR/dP$ as a function of $R$.

**Figure 4: Analysis of electrical heating. a**, **b**, **c**, Temperature dependence of the resistance $R$, derivative of resistance with power $dR/dP$ and heat resistance $R^h$. Symbols are experimental data, and red overlapping lines are power law fits. **d**, Heat resistance diagram of the device.

**Figure 5: Response speed of DGBLG hot electron bolometer.** Normalized photo response from pump probe laser pulses as a function of delay $t_D$ at 4.55 K and 10 K with bias $I_{dc}$ = 185 nA. The sample is gated to charge neutrality with a displacement field of 0.45V/nm. Lines show best fits assuming exponential decay of hot electron temperature $T_e$ with a thermal response time $\tau$ and considering nonlinear $T_e$ dependence of graphene resistance. Solid line (red) and dot-dashed line (blue) are for $\tau$ = 0.25 ns and 0.1 ns, respectively.

Fig.1

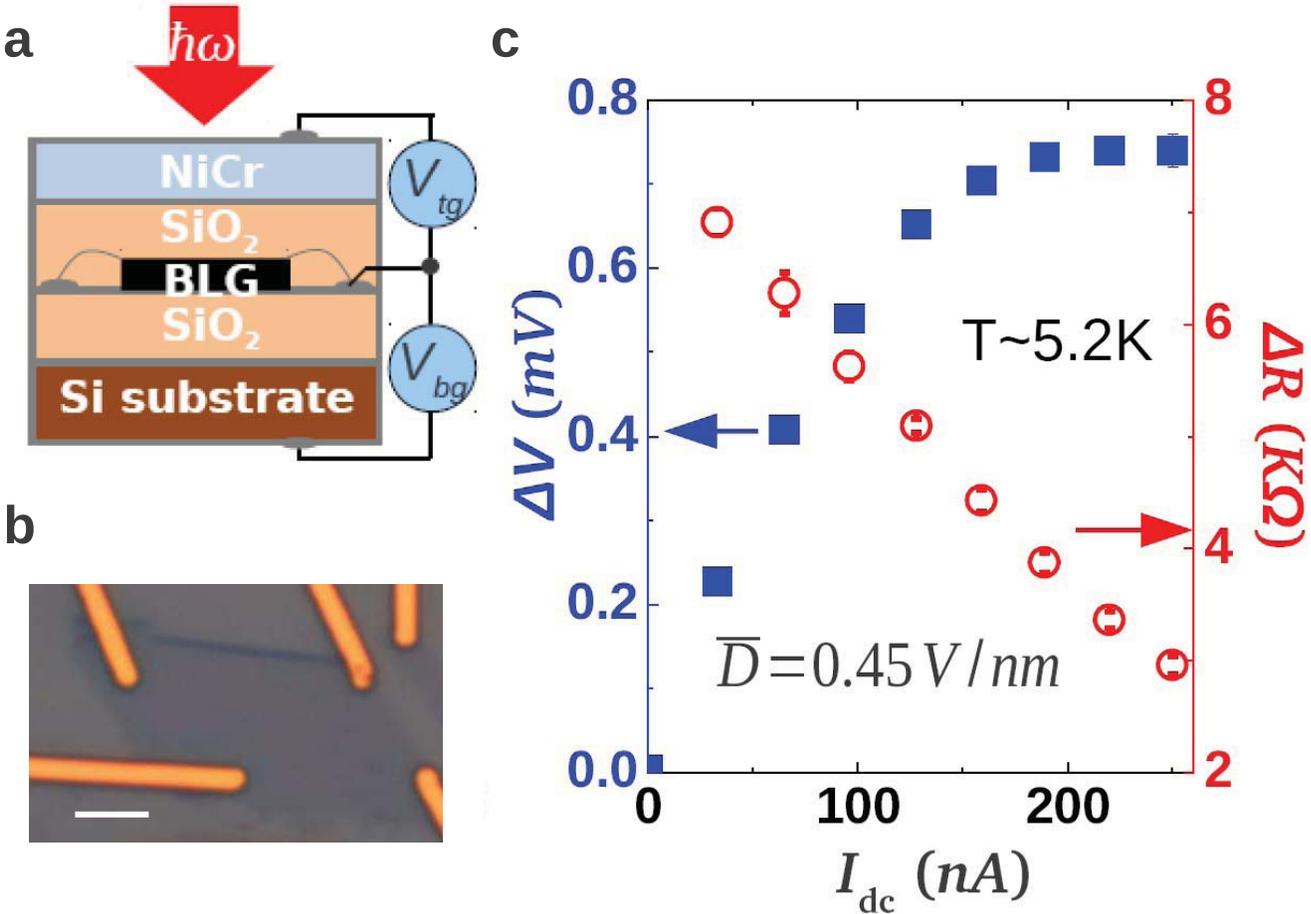

**Fig.2**

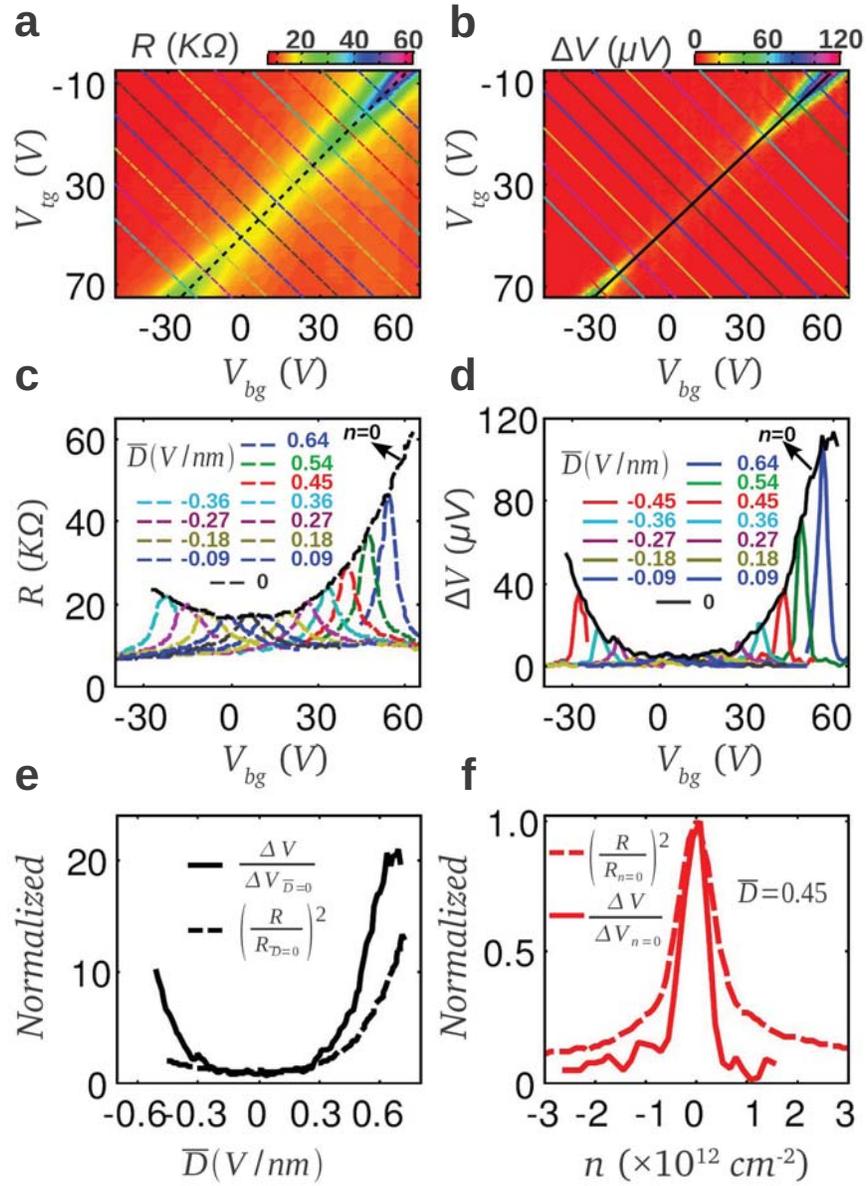

**Fig.3**

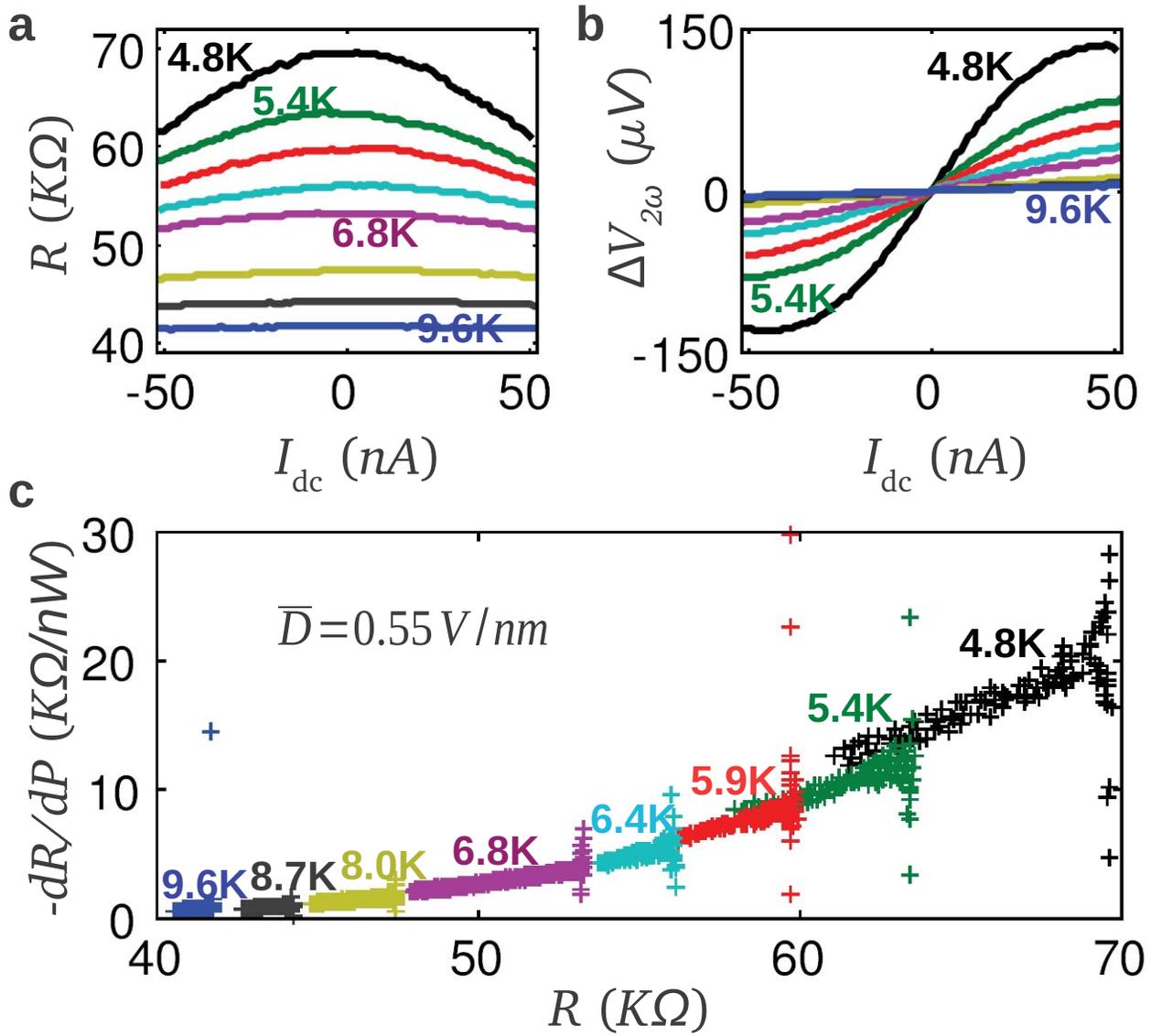

Fig.4

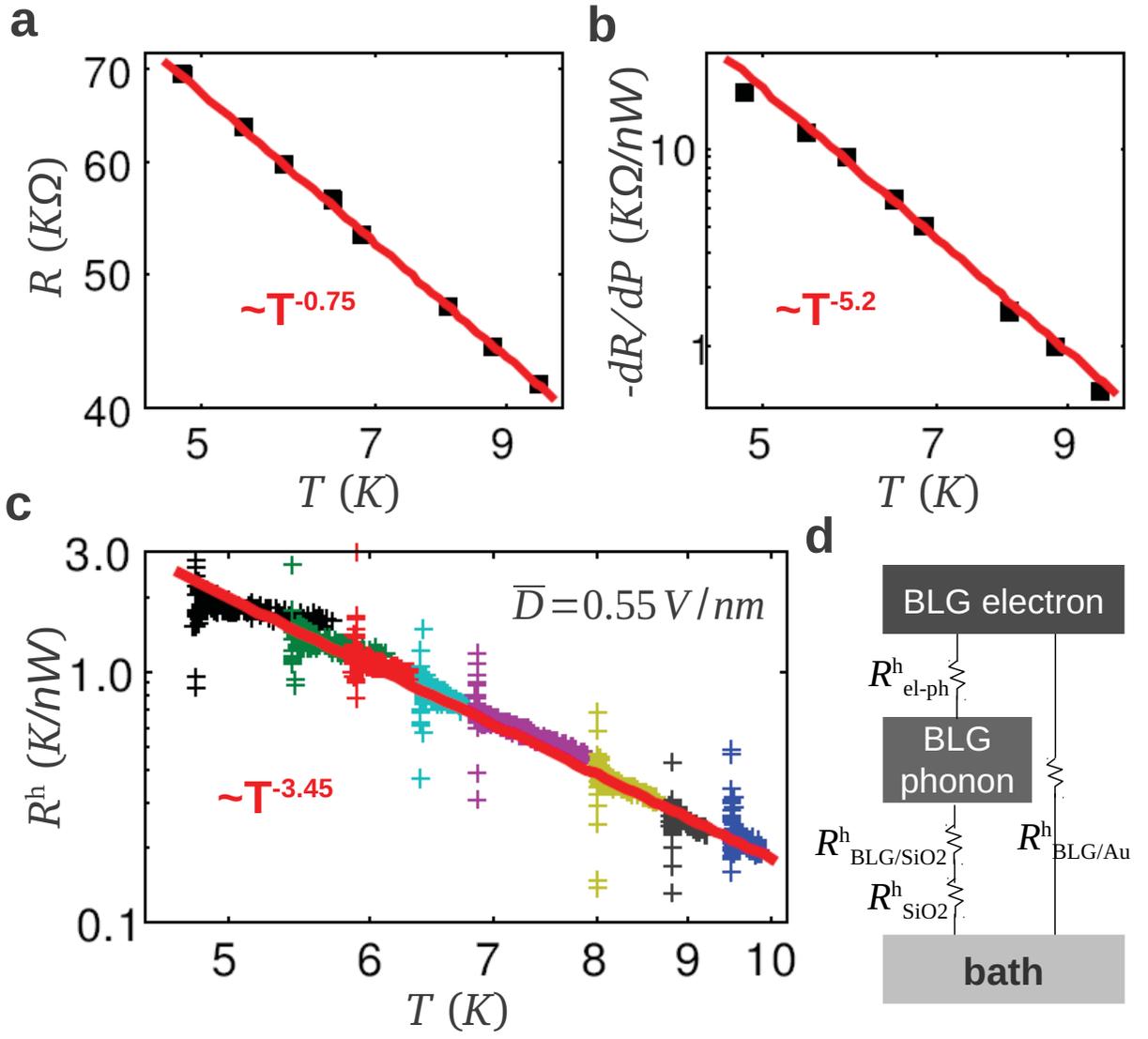

Fig.5

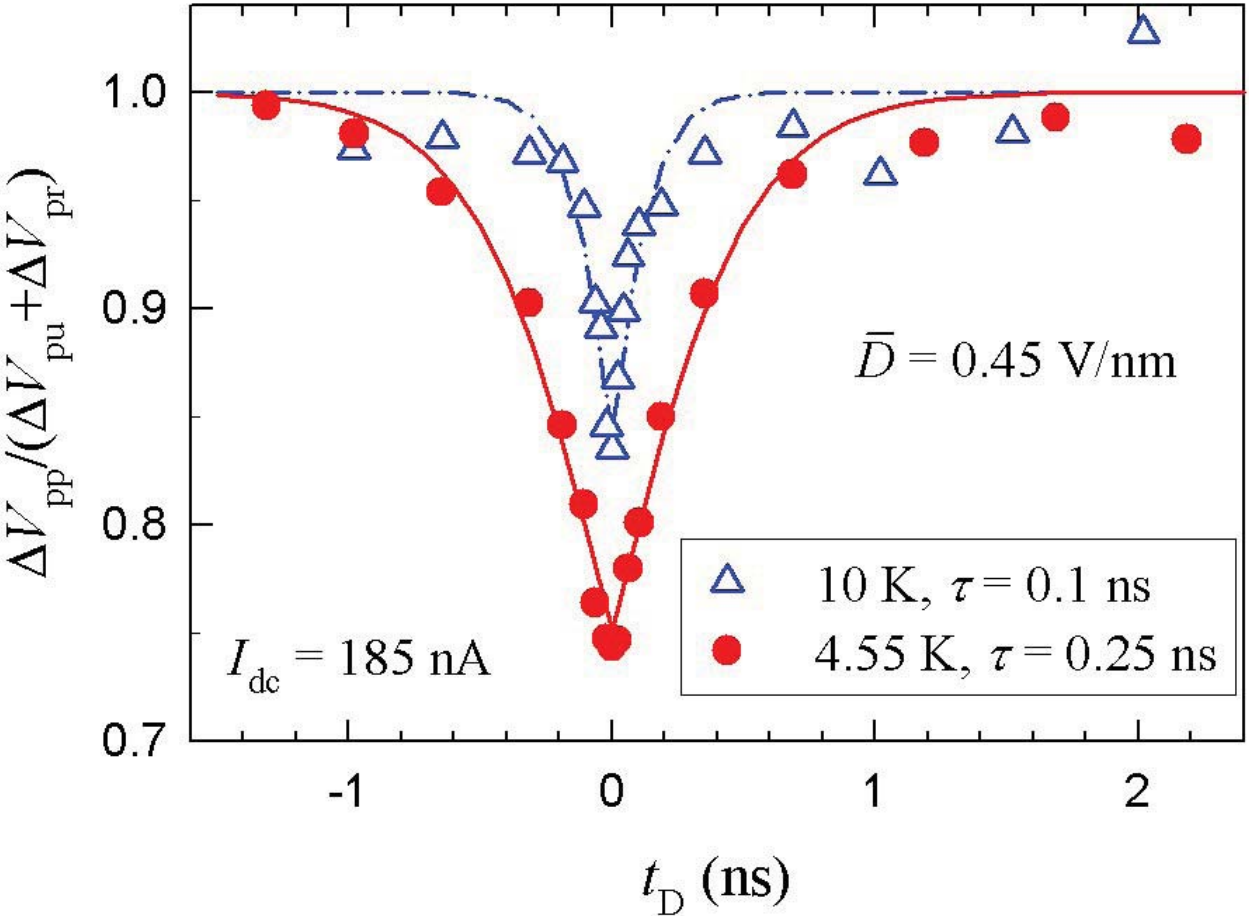